\documentclass[11pt,amssymb,preprint,showpacs]{revtex4}
\usepackage{hyperref}
\usepackage{amsmath}
\usepackage{graphicx}
\usepackage{amssymb} 
\usepackage{amsfonts} 

\begin{document}


\title{Quantum Mechanics and Closed Timelike Curves}
\author{Florin Moldoveanu}
\affiliation{Department of Theoretical Physics, National Institute for Physics and Nuclear Engineering, P.O. Box M.G.-6, Magurele, Bucharest, Romania}
\email{fmoldove@gmail.com}

\begin{abstract}
General relativity allows solutions exhibiting closed timelike curves. Time travel generates paradoxes and quantum mechanics generalizations were proposed to solve those paradoxes. The implications of self-consistent interactions on acausal region of space-time are investigated. If the correspondence principle is true, then all generalizations of quantum mechanics on acausal manifolds are not renormalizable. Therefore quantum mechanics can only be defined on global hyperbolic manifolds and all general relativity solutions exhibiting time travel are unphysical. 
\end{abstract}

\pacs{04.20.Cv, 03.30.+p, 02.10.Ab}

\maketitle

\section{Introduction}
Quantum mechanics and general relativity are core physics theories. Intense research is performed today to obtain a unified theory which will address the current incompatibilities between the two theories. It is conceivable that one or both of the theories would have to be generalized in order to obtain a coherent theory of nature. This paper attempts to eliminate an entire class of potential quantum mechanics generalizations that were proposed to address closed timelike curves (CTC's), or ``time machines" which occur naturally in general relativity. 
 
\section{ General relativity and time travel } 
In special relativity causality holds due to the hyperbolic nature of the Minkowski metric that clearly separates past from future. In general relativity however, because the Einstein equations are local equations and space-time can be curved by matter, one can construct solutions that exhibit closed timelike curves on a global scale.

There are a surprisingly large number of such solutions \cite{Visser1}, and here is only a partial list: van Stockum spacetime \cite{ Stockum1}, G\"{o}del's rotating universe \cite{ Godel2}, Tipler's cylinders \cite{Tipler1}, Kerr geometries \cite{Kerr1}, Wheeler wormhole \cite{ Wheeler1}, Morris-Thorne traversable wormholes \cite{Morris1}, Gott's infinite cosmic strings \cite{Gott1}, and Alcubierre's ``warp drive" spacetime \cite{Alcubierre1}. Thus CTCs cannot be easily discarded as unphysical solutions. It is not hard to imagine paradoxes created by time travel. Let us start by reviewing the usual problems of closed causal loops. There are only two known classes of time travel paradoxes: the grandfather paradox, and the creation of information from nothing.

In the grandfather paradox, a time traveler goes back in time and prevents his grandfather to meet his grandmother, thus preventing his own birth. In the information paradox, a person is handed the blue-prints of a time machine by its older self, constructs the machine and use it to hand himself the blue-prints. Who invented the blue-prints?

\subsubsection{The grandfather paradox}
The grandfather paradox is usually encountered in classical mechanics. A typical philosophical argument against CTCs is ``free will": when I go back in time it is my free will to kill my younger self.

The ``free-will" counter-argument is as follows: it is my free will to walk on the ceiling, but the laws of physics prevent it and in the same way when I want to complete an inconsistent CTC the laws of physics would prevent it no matter how hard I try. All I can achieve is a consistent CTC \cite{Novikov2}.

Since ``free-will" is a fuzzy philosophical concept, to make the problem mathematically tractable, it is usual to consider the collision of billiard balls at the mouth of a wormhole.

A well known general relativity solution exhibiting time travel is the wormhole solution. A spherically symmetric and static traversable wormhole is represented by the following spacetime metric: 
\begin{equation}
ds^2 = -e^{2\Psi (r)} dt^2 + \frac{dr^2}{1-b(r)/r} + r^2 (d\theta^2 + \sin^2 \theta d\phi^2)
\end{equation}
where $\Psi(r)$ and $b(r)$ are arbitrary functions of the radial coordinate, $r$ \cite{Morris2}. 
Acceleration of one of the wormhole mouths can introduce a time delay which transforms the wormhole into a time machine.

In this case we have to address the ``Polchinski paradox"\cite{ Friedman1}. Consider that a billiard ball falls through a wormhole, travels back in time, and collides with its younger self, preventing it to fall in the wormhole in the first place. Similar paradoxes have been obtained by Rama and Sen\cite{Rama1} when they considered collisions of objects of different mass.

\begin{figure}[ht]
\centering \leavevmode
\includegraphics[width=6.5cm]{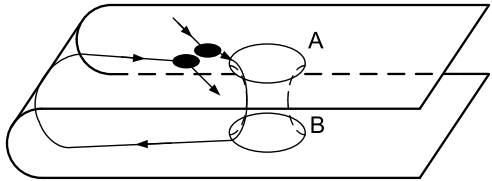}
\caption{ Self-consistent collision of a billiard ball with itself in the presence of a traversable wormhole. This is an embedding diagram for a wormhole connecting two regions of flat Minkowski space. There is a time differential between the two mouths A and B allowing the billiard ball to arrive back in time and collide with its younger self.} \label{Figure6}
\end{figure}

There is only one way to avoid those paradoxes: eliminate the initial conditions that can lead to them\cite{Rama1}. A self consistent collision would look like the interaction shown in Fig. \ref{Figure6}.

Restricting initial conditions is very disturbing, because no such mechanism is known in the macroscopic world. Just the mere presence of a CTC region in our causal future would have consequences here and now. We may not be visited by time travelers from the future because we are uninteresting, but nobody has yet observed any evidence of restricted initial conditions. Elementary particles can be accelerated very precisely in any direction one wants.

\subsubsection{Time travel and quantum mechanics}
Originally it was suggested that quantum mechanics may play a role in achieving the selection of the appropriate initial conditions \cite{Friedman1} and cure the classical multiplicity of solutions (which may also include inconsistent collisions). For example, by performing a sum-over-consistent-histories in a WKB approximation we may obtain only the correct and unique initial condition corresponding to the consistent self-interaction. 

It is true that a final theory of quantum mechanics in curved space-time does not exist yet, and maybe close to the wormhole mouth new physical laws may be at play. The problem is that the world line of the billiard ball originates far away from the wormhole in flat space-time where regular physics is applicable. 

Now, however unlikely, assume that such restrictions are somehow in place, due to a future, yet undiscovered, unified theory. Are CTCs compatible with quantum mechanics?

First, the grandfather paradox does not occur in quantum mechanics because in quantum mechanics one encounters qbits. The classical contradiction could coexist in a superposition of quantum states\cite{Svozil1}, and the only way to get a real contradiction is if summing over all states the total probability is no longer one. Therefore the contradiction would manifest itself as a unitarity problem \cite{Goldwirth1}. Propagating along a CTC, the wavefunction can become causality trapped. When the probability to detect the particle outside the CTC region is more than one, this corresponds to the paradox of creating information from nothing. When the probability is smaller than one, then part of the original physical system becomes causally trapped inside the CTC region in a cyclical history that repeats itself forever.

But the problem is even worse. Quantum mechanics as it is known is incompatible with CTCs as pointed out by Jacobson\cite{Jacobson1}: there are ambiguities in computing expectation values. This is expected if one has non-unitary evolution since the Cauchy problem is ill defined.

Several possible generalizations for quantum mechanics were proposed by Hartle\cite{Hartle1}, Anderson\cite{Andersen1}, and Fewster\cite{Fewster1}. One may argue that those generalizations may have objectionable feature like the fact that the presence of a CTC region in our causal future can imply measurable departures from today's predictions of standard quantum mechanics. We have already encountered those features in the classical physics in the form of restriction to initial conditions, and the fact the free will may only be an illusion. If the CTC region is well ahead in our causal future, then the measurable effects of those departures from standard quantum mechanics and the usual classical physics may be negligible. To be able to reject those generalizations we need to investigate additional consequences of those theories.

For example Hawking showed that quantum coherence is lost\cite{Hawking3} in a CTC and therefore one cannot gain any information from time travelers from the future. But while time travel may be of no value to gain the knowledge of the future, this is no argument to reject time travel altogether. 

Also we can construct non-contradictory quantum field theories on curved space-time only on global hyperbolic spaces\cite{Kay1}. Again, the lack of our current knowledge does not constitute a proof that a future, yet undiscovered, unified theory of general relativity and quantum mechanics may not be able to provide the correct generalization of quantum mechanics and the justification of the initial condition selection.

There is yet another reason to believe that time travel is impossible as pointed by 
Hawking in his ``Chronology Protection Conjecture''\cite{Hawking2}. The vacuum polarization effects will get amplified by a CTC region resulting in a gravitational back-reaction that will destroy the CTC region. Since we lack a unified theory, we cannot just base the rejection of time travel on semi-classical approximations. As a counter example Visser showed that a Roman ring of wormholes can create a time machine without stability problems \cite{Visser2}.

A way to solve the paradoxes of time travel is to demand a global self consistency condition that will guarantee that all self-interactions are consistent \cite{Novikov2}. The laws of physics are usually defined locally and the fundamental reason for this is that on a curved manifold, the tangent space is defined at each point. Local laws seem to prevent the existence of any global constraints. 

There are two possible counter arguments to the local physical laws argument. First, quantum mechanics is non-local. Second, consider the motion in phase space or that of an incompressible fluid. In this case closed trajectories do form and there are no local contradictions. This is actually a very intuitive way to picture what should happen on CTC region that is globally consistent.

For the first counter-argument, we will investigate below the consequences of a global self-consistency principle and find that it leads to non-renormalizable theories. For the second counter-argument, one cannot define a metric on a phase space. There is much more local freedom on a metric manifold, than on a symplectic manifold. On a symplectic manifold one can define only a skew-symmetric bilinear form and have only global invariants. In phase space one cannot have arbitrary changes of coordinates, because that would violate Hamilton's equations. 

\subsubsection{ The Consequences of Consistent Closed Timelike Curves }
Let us consider more and more violent collisions of the billiard ball in the Polchinski-type collision. At some point, the billiard balls will break and generate a paradox. Equivalently, one can consider repeating the consistent self-interaction with the same initial conditions, but with more and more brittle billiard balls of the same mass and shape.

One can even consider manufacturing billiard balls made of $2, 3, \dots, n$ pieces , with a small explosive in the middle that will break the ball in its constituent pieces during any collision.

The key point is that one repeats the experiment with all those balls, preserving the same initial conditions corresponding to a self-consistent interaction. 

With those preparations, let us assume that the billiard ball is made of two equal halves ($L$ and $R$) that are loosely connected (by a weak material) and will separate if during the collision the momentum transfer in the center of mass referential system exceeds a particular threshold value.

Now repeat the collision experiment with the same initial condition but with increased brittleness (reduced threshold for separation) of the billiard balls of the same shape and mass. From the point of view of a global self-consistency principle, nothing is changed.

In one of the repeated experiments, at some point the momentum transfer is going to exceed the threshold value in a center of mass coordinate system and the billiard ball will break. One can enforce the pieces separation after breaking for example by adding a positive electric charge to each piece. After the collision, there are three possible outcomes:
\begin{enumerate}
\item No piece enters the wormhole,
\item Only one piece enters the wormhole and causes the earlier collision,
\item The two pieces $L$ and $R$ remain together and follow same self-consistent trajectory.
\end{enumerate}

If no piece enters the wormhole, then the earlier collision did not take place and we have a paradox. If only one piece causes the collision, because it's mass is half of the original billiard ball, the momentum transfer falls below the separation threshold, and no longer causes the breaking, resulting in another paradox. Only if the billiard ball pieces stay together are all paradoxes avoided. This implies that the self-consistency principle requires infinite strength to maintain cohesion for \emph{all} billiard balls participating in a self-consistent interaction and this is just not true. The global consistency condition imposes an impossible unphysical demand on the local physics of collision which can also happen far away from the wormhole mouths where standard physics is applicable.

When indestructible elementary particles are used instead of billiard balls, the contradictions are not avoided because at high enough energies other particles are going to be generated. As other particles are generated, the energy of the original particle is reduced below the generation threshold thus preventing the generation in the first place and again a contradiction ensues.

\emph{If even in classical physics one encounters infinities, wherever the correspondence principle is valid there is no hope to obtain a renormalizable quantum theory in curved space-time containing CTCs.}

The only possible escape of the conclusion is if the correspondence principle may not hold and classical physics can not emerge from the quantum world. This may be the case at Plank scale where the regular space time manifold may cease to be well defined. In this range we do need a unified theory of general relativity and quantum mechanics.

\section{Conclusion}

We can now conclude that closed causal loops are forbidden in general in nature in the range of the validity of the correspondence principle. Quantum mechanics can only be defined on global hyperbolic manifolds and all general relativity solutions exhibiting time travel are unphysical. As Hawking put it, the world is indeed safe for historians.

\end{document}